\newcommand\submitms{n}

\documentclass[12pt]{article}

\pdfoutput=1
\usepackage{scicite}

\usepackage{astjnlabbrev-nature}
\usepackage{graphicx}
\usepackage{caption}
\usepackage{color}
\usepackage{movie15}
\usepackage[english]{babel}

\usepackage{times}

\newcommand\degree{\mbox{$^\circ$}}
\newcommand\degrees{\degree}
\DeclareSymbolFont{UPM}{U}{eur}{m}{n}
\DeclareMathSymbol{\umu}{0}{UPM}{"16}
\let\oldumu=\umu
\renewcommand\umu{\ifmmode\oldumu\else$\oldumu$\fi}
\newcommand\micro{\umu}
\newcommand\micron{\micro m}
\newcommand\microns{\micro m}
\newcommand\arcsec[0]{$^{\prime\prime}$}
\let\oldsim=\sim
\renewcommand\sim{\ifmmode\oldsim\else$\oldsim$\fi}
\let\oldpm=\pm
\renewcommand\pm{\ifmmode\oldpm\else$\oldpm$\fi}
\let\oldmsp=\sp
\let\oldmsb=\sb
\def\sp#1{\ifmmode
           \oldmsp{#1}%
         \else\strut\raise.85ex\hbox{\scriptsize #1}\fi}
\def\sb#1{\ifmmode
           \oldmsb{#1}%
         \else\strut\raise-.54ex\hbox{\scriptsize #1}\fi}

\newenvironment{sciabstract}{%
\begin{quote} \bf}
{\end{quote}}

\newcounter{lastnote}

\title{Thermal structure of an exoplanet atmosphere from phase-resolved emission spectroscopy}

\author
{Kevin B. Stevenson$^{1,13\ast}$, Jean-Michel D\'esert$^2$, Michael R. Line$^{3}$, Jacob L. Bean$^1$,\\
 Jonathan J. Fortney$^{3}$, Adam P. Showman$^{4}$, Tiffany Kataria$^{4}$, Laura Kreidberg$^1$,\\
 Peter~R.~McCullough$^{5,6}$, Gregory W. Henry$^{7}$, David Charbonneau$^{8}$, Adam Burrows$^{9}$,\\
 Sara Seager$^{10}$, Nikku Madhusudhan$^{11}$, Michael H. Williamson$^{7}$ \& Derek Homeier$^{12}$\\
\vspace*{-6 pt}
\\
\normalsize{$^{1}$Department of Astronomy and Astrophysics, University of Chicago, }\\
\normalsize{5640 S Ellis Ave, Chicago, IL 60637, USA}\\
\normalsize{$^{2}$CASA, Department of Astrophysical and Planetary Sciences, University of Colorado, }\\
\normalsize{389-UCB, Boulder, CO 80309, USA}\\
\normalsize{$^{3}$Department of Astronomy and Astrophysics, University of California, }\\
\normalsize{Santa Cruz, CA 95064, USA}\\
\normalsize{$^{4}$Department of Planetary Sciences and Lunar and Planetary Laboratory, }\\
\normalsize{The University of Arizona, Tucson, AZ 85721, USA}\\
\normalsize{$^{5}$Space Telescope Science Institute, Baltimore, MD 21218, USA}\\
\normalsize{$^{6}$Department of Physics and Astronomy, Johns Hopkins University, }\\
\normalsize{3400 North Charles Street, Baltimore, MD 21218, USA}\\
\normalsize{$^{7}$Center for Excellence in Information Systems, Tennessee State University, }\\
\normalsize{Nashville, TN 37209, USA}\\
\normalsize{$^{8}$Department of Astronomy, Harvard University, Cambridge, MA 02138, USA}\\
\normalsize{$^{9}$Department of Astrophysical Sciences, Princeton University, Princeton, NJ 08544, USA}\\
\normalsize{$^{10}$Dept. of Earth, Atmospheric, and Planetary Sciences, Dept. of Physics,}\\
\normalsize{ Massachusetts Institute of Technology, 54-1718, 77 Mass. Ave., Cambridge, MA  02139, USA}\\
\normalsize{$^{11}$Institute of Astronomy, University of Cambridge, Cambridge CB3 OHA, UK}\\
\normalsize{$^{12}$Centre de Recherche Astrophysique de Lyon, UMR 5574, CNRS, Universit\'e de Lyon,}\\
\normalsize{ \'Ecole Normale Sup\'erieure de Lyon, 46 All\'ee d'Italie, F-69364 Lyon Cedex 07, France}\\
\normalsize{$^{13}$NASA Sagan Fellow}\\
\vspace*{-6 pt}
\\
\normalsize{$^\ast$To whom correspondence should be addressed; E-mail:  kbs@uchicago.edu.}
}

\date{}

\begin{document} 

\baselineskip24pt

\maketitle 

\begin{sciabstract}
Exoplanets that orbit close to their host stars are much more highly irradiated than their Solar System counterparts.   Understanding the thermal structures and appearances of these planets requires investigating how their atmospheres respond to such extreme stellar forcing.  We present spectroscopic thermal emission measurements as a function of orbital phase (``phase-curve observations'') for the highly-irradiated exoplanet WASP-43b spanning three full planet rotations using the {\em Hubble Space Telescope}.  With these data, we construct a map of the planet's atmospheric thermal structure, from which we find large day-night temperature variations at all measured altitudes and a monotonically decreasing temperature with pressure at all longitudes.  We also derive a Bond albedo of $0.18^{+0.07}_{-0.12}$ and an altitude dependence in the hot-spot offset relative to the substellar point.
\end{sciabstract}

Previous exoplanet phase-curve observations\cite{Knutson2007,Knutson2009,Crossfield2010,Cowan2012,Knutson2012,Lewis2013,Maxted2013} have revealed day-night temperature contrasts and hot-spot offsets relative to the substellar point (the point at which the host star would be perceived to be directly overhead).  However, these observations were limited to broadband photometry; therefore, the altitudes probed by the phase curves were not uniquely constrained.  Spectroscopic phase curves can break previous degeneracies by permitting us to uniquely identify the main atmospheric opacity source within the observed bandpass and infer the planet's atmospheric temperature-pressure profile as a function of orbital phase\cite{Seager1998,Burrows2000,Guillot2002,Sudarsky2003,Fortney2006b}.

The WASP-43 system contains a transiting Jupiter-size exoplanet on a 19.5-hour orbit around its K7 host star\cite{Hellier2011}.  Previous measurements\cite{Gillon2012,Wang2013,Chen2014,Blecic2014} of its dayside thermal emission detect no signs of a thermal inversion and suggest low day-night energy redistribution.  However, the precise thermal structure of the dayside atmosphere remains unknown without higher resolution observations, and the planet's global energy budget and atmospheric heat-redistribution efficiency is poorly constrained without observations of the nightside.

Over 4 - 7 November 2013, we used the Wide Field Camera 3 (WFC3) instrument from the {\em Hubble Space Telescope} ({\em HST}) to observe three nearly-consecutive orbits of WASP-43b.  The planet orbits so close to its host star that it is tidally locked. Therefore, orbital phase is equivalent to rotational phase for the planet, and observations over a complete orbit allow us to map the entire surface of the planet.  {\em HST} also acquired data for three primary transits and two secondary eclipses, where the planet passes in front of and behind its host star, respectively, between 9 November 2013 and 5 December 2013. All of the observations used the G141 grism (1.1 -- 1.7 {\microns}) and the bi-directional spatial scan mode. 

\begin{figure}
\centering
\if\submitms n
    \includegraphics[width=16cm]{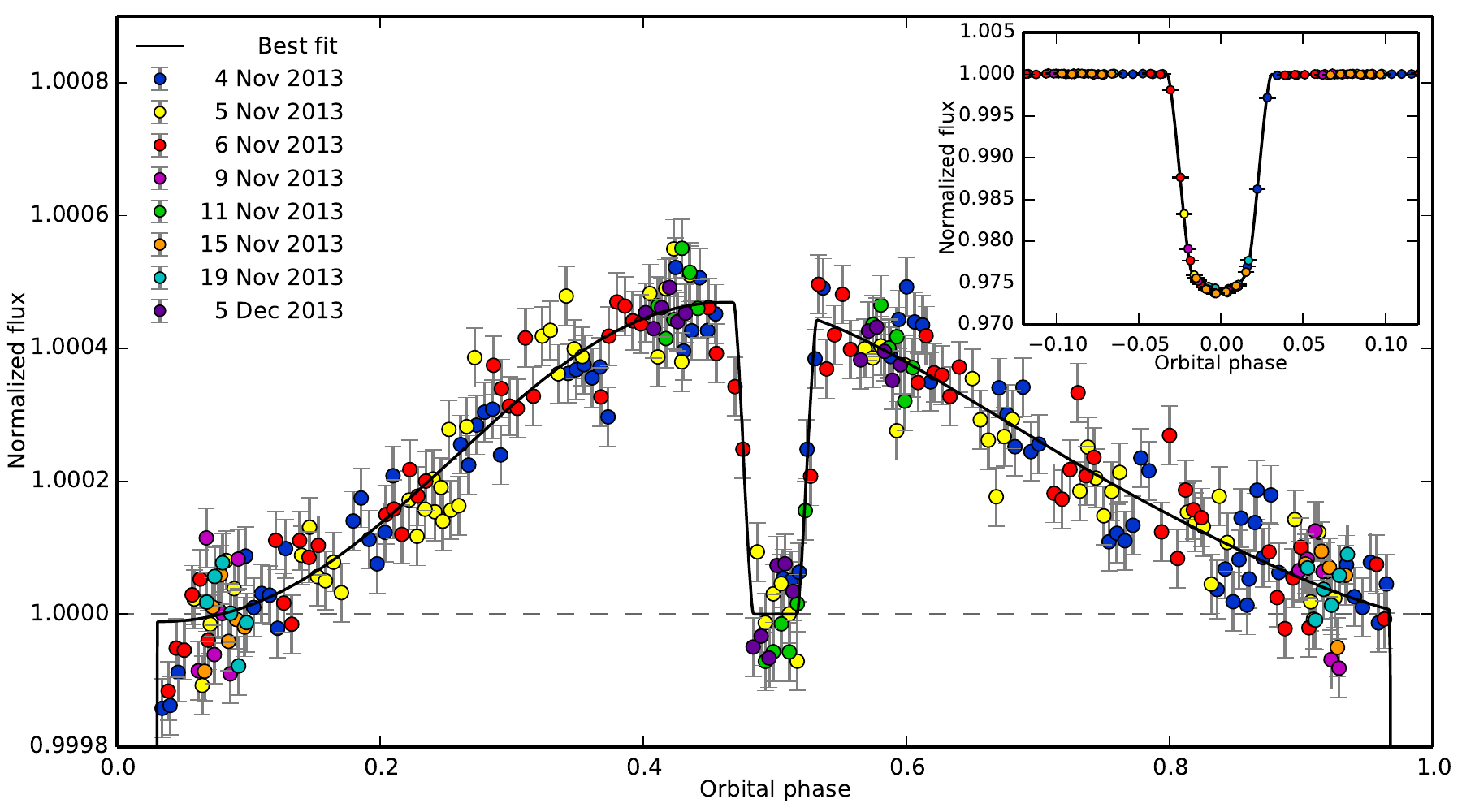}
\fi
\caption{\label{fig:whitelc}
{\bf Band-integrated phase curve of WASP-43b.}  
The systematics-corrected flux values are binned in time, normalized to the stellar flux, and have 1$\sigma$ error bars.  Each color represents data acquired from a different {\em HST} visit.  The phase curve depicts steadily increasing and decreasing observed flux which originates from different longitudes of the tidally-locked planet as it makes one complete rotation.  Light from the planet is blocked near an orbital phase of 0.5 as it is eclipsed by its host star.  The model phase curve maximum occurs 40 {\pm} 3 minutes prior to the midpoint of secondary eclipse, which corresponds to a shift of 12.3 {\pm} 1.0$\degrees$ East of the substellar point.  The model phase curve minimum occurs 34 {\pm} 5 minutes after the primary transit midpoint, or 10.6 {\pm} 1.4$\degrees$ West of the anti-stellar point.  As a result, maximum planetary emission occurs 0.436 {\pm} 0.005 orbits after the observed minimum (for depths probed by these observations) and the shape of the phase curve is asymmetric.  Inset, for comparison, is the white light curve primary transit.  It is interesting to note that the observed flux values are consistently low for $\sim$30 minutes after transit egress.
}
\end{figure}

Using custom software\cite{Stevenson2013,SI2014}, we reduced the data and extracted the spectra.  We produced time-series spectroscopy by dividing the spectra into 15 0.035-{\micron}-wide channels (7 pixels, resolution $R = \lambda / \Delta\lambda \sim 37$).  We also produced band-integrated ``white'' light curves to resolve finer details in the shape of the phase curve (Fig. \ref{fig:whitelc}).  We simultaneously fit the light curves using transit and uniform-source eclipse models\cite{Mandel2002}, a baseline flux for each {\em HST} scan direction, two standard model components for {\em HST} orbit-long and visit-long systematics, and a sinusoidal function to represent the phase variation \cite{Kreidberg2014,SI2014}.  We estimate uncertainties using a differential-evolution Markov-chain Monte Carlo (DE-MCMC) algorithm\cite{Stevenson2013} and utilize an independent analysis pipeline\cite{Kreidberg2014} to confirm our light-curve fits. 

The white light phase curve (Fig. \ref{fig:whitelc}) reveals a distinct increase in flux as the tidally-locked dayside rotates into view.  The flux peaks prior to secondary eclipse (eastward of the substellar point) and then decreases until the planet transits in front of its host star.  Because the phase curve minimum occurs west of the anti-stellar point, we detect a strong asymmetry ($\sim10\sigma$) in the shape of the observed phase curve.  We measure a white light curve eclipse depth that is consistent with the peak-to-peak planet flux variation.  This confirms a relatively cool night side and poor heat redistribution.  Table S\ref{tab:WhiteData} lists our best-fit parameters with uncertainties.

\begin{figure}
\centering
\if\submitms n
    \includegraphics[width=15cm]{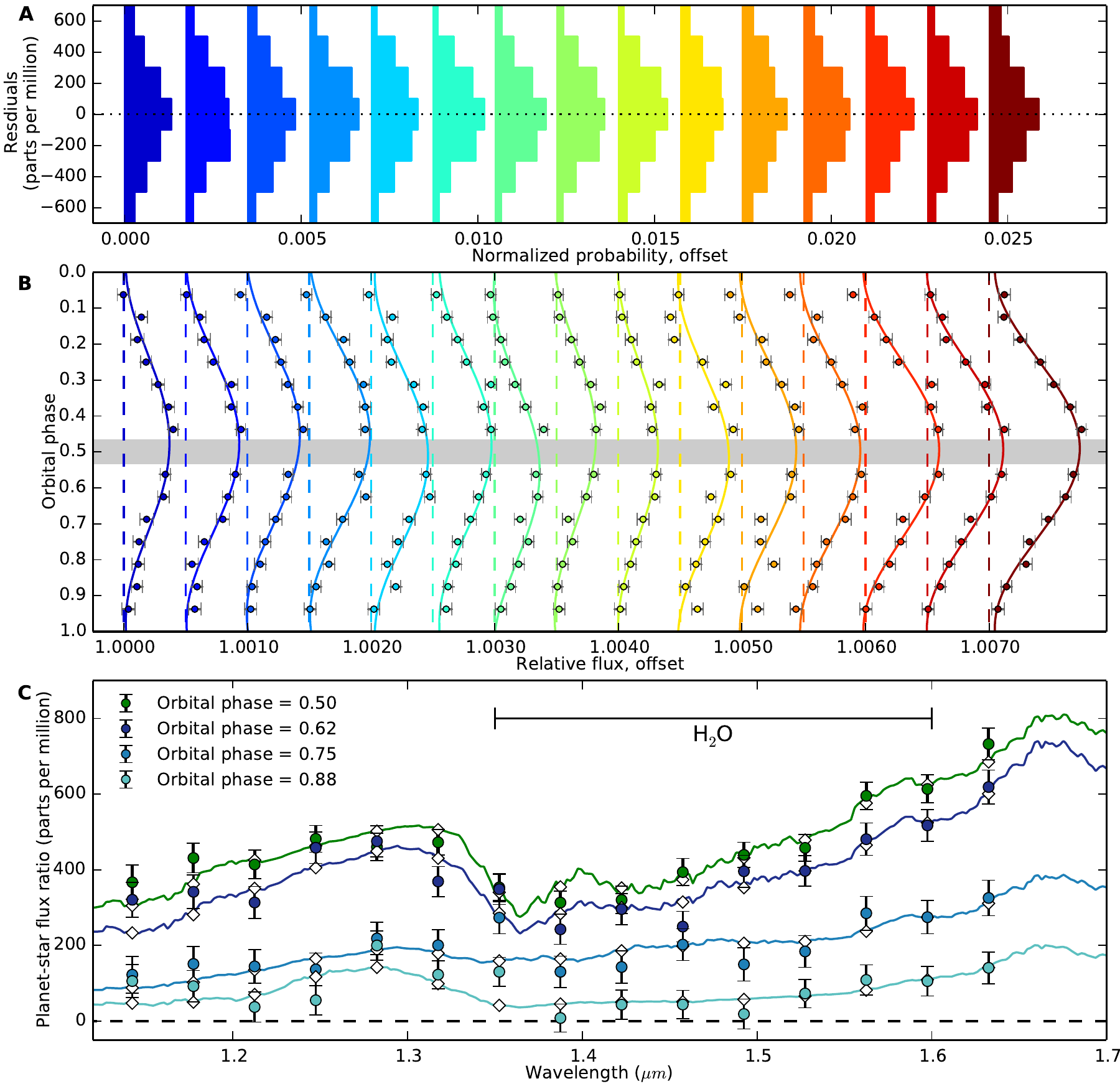}
\fi
\caption{\label{fig:1DEmSpec}
{\bf Phase-resolved emission spectrum of WASP-43b relative to the stellar flux.}
{\bf A}, The histograms of the unbinned phase-curve residuals are separated horizontally by wavelength (colors, defined on the abscissa of panel {\bf C}) for clarity.  The residuals are Gaussian distributed with a zero mean and show no evidence of correlated noise.  
{\bf B}, We show binned phase curves (colored points with 1$\sigma$ error bars) and best-fit models (colored lines).  The planet emission is normalized with respect to the stellar flux and separated horizontally by wavelength for clarity.  The gray region depicts the time of secondary eclipse.  
{\bf C}, We illustrate a subset of data points from panel {\bf B}, except plotted as a function of wavelength and with best-fit atmospheric models (colored lines).  White diamonds depict the models binned to the resolution of the data.  For clarity, we only display planet-to-star flux ratios at four planet phases: full (0.5, secondary eclipse), wanning gibbous (0.62), half (0.75), and wanning crescent (0.88).  In Figs. S\ref{fig:PCall} -- S\ref{fig:EmSpecall}, we provide full 1D and 2D representations of panels {\bf B} and {\bf C}.  A time-lapse video of the planet's phase-resolved emission spectrum is available in Movie S1.
}
\end{figure}

We gain additional information by decomposing the white light phase curve into 15 spectrophotometric channels (Fig. \ref{fig:1DEmSpec}).  The spectrally-resolved phase curves exhibit wavelength-dependent amplitudes, phase shifts, and eclipse depths (Table S\ref{tab:SpecData}).  We use the measured phase-resolved emission spectra (Fig. \ref{fig:1DEmSpec}C) to infer the temperature structure and molecular abundances at 15 binned orbital phases (each of width 0.0625).  We fit atmospheric models to these spectra using a DE-MCMC approach from the CHIMERA Bayesian retrieval suite\cite{Line2014-C/O}.  For each phase, a five-parameter, double-gray radiative equilibrium solution parameterizes the planet's temperature structure\cite{Parmentier2013}.  The models include six thermochemically plausible and spectrally prominent absorbers (H\sb{2}O, CH\sb{4}, CO, CO\sb{2}, NH\sb{3}, and H\sb{2}S).  We find that water is the only absorber to significantly influence the phase-resolved emission spectra \cite[{\rm Fig. \ref{fig:1DEmSpec}}]{Kreidberg2014b}.  The model spectra are in good agreement with the data, achieving a typical $\chi^2$ value of 18 with 15 data points and 6 relevant free parameters (Fig. S\ref{fig:EmSpecall}).

Using the atmospheric models to estimate the day- and night-side fluxes, we find that the planet redistributes heat poorly\cite[$\mathcal{F} = 0.503^{+0.021}_{-0.003}$, {\rm where} $\mathcal{F} = 0.5 \rightarrow 1$ {\rm spans the range from zero to full heat redistribution}]{SI2014}.  This is predicted to occur when the radiative timescale is shorter than the relevant dynamical timescales, including those for wave propagation and advection over a hemisphere\cite{Perez-Becker2013}.  Poor redistribution has been inferred before, but only for hot Jupiters receiving significantly greater stellar flux than WASP-43b\cite{Cowan2012,Maxted2013}.  We estimate the fraction of incident stellar light reflected by WASP-43b's atmosphere by computing the day- and night-side bolometric fluxes from the model spectra and find a Bond albedo of $0.18^{+0.07}_{-0.12}$. This method assumes energy balance with the parent star but requires no detection of reflected light\cite{SI2014}.  The low Bond albedo is consistent with model predictions that hot Jupiters absorb most of the flux incident upon them\cite{Marley1999,Sudarsky2003,Burrows2008}. 

\begin{figure}
\centering
\if\submitms n
    \includegraphics[width=11.0cm]{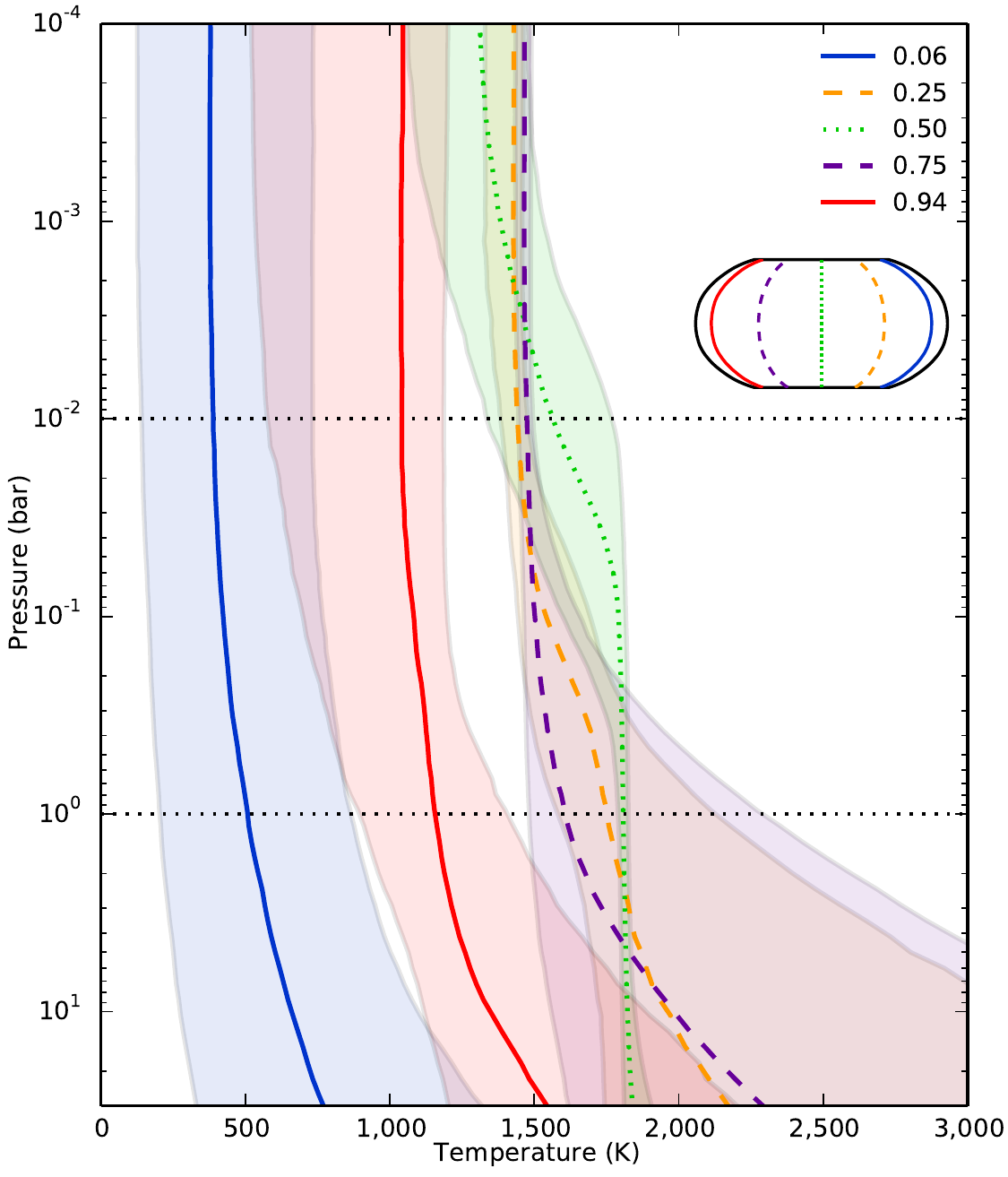}
\fi
\caption{\label{fig:tp}
{\bf Thermal profiles of WASP-43b at select orbital phases.}
Higher pressures indicate deeper within the planet's atmosphere.  Colored curves depict median values with 1$\sigma$ uncertainty regions for the assumed parameterization of the retrieval.  We illustrate the temperature asymmetry on the planet's night side immediately before and after transit (orbital phase = 0.0625 and 0.9375), the similar thermal profiles on WASP-43b's morning and evening terminators (0.25 and 0.75), and the dayside-averaged profile (0.5).  The {\em HST}/WFC3 measurements probe the atmosphere primarily between 0.01 and 1.0 bar (horizontal dotted lines).  The retrieved model profiles are 1D representations of the disk-integrated flux values at each phase.  However, because the emitted flux values at these wavelengths are near the peak of the Planck curve, the flux goes as $T^{5}$ or more and the disk-integrated thermal profiles are heavily weighted towards the hotter dayside.  As a result, there is no significant change in the modeled temperature structure over half of the orbital phases (0.25 $\rightarrow$ 0.75, when the substellar point is visible).  We plot individual pressure-temperature profiles with 1$\sigma$ uncertainty regions in Fig. S\ref{fig:tpall}.  A time-lapse video of WASP-43b's phase-resolved thermal profile is available in Movie S1.
}
\end{figure}

The atmospheric model fits reveal information about WASP-43b's phase-dependent thermal structure at the pressure levels probed by these observations (Fig. \ref{fig:tp}).  Depending on the wavelength and phase, these pressures range from 0.01 to 1 bar (Fig. S\ref{fig:tpall}).  The retrieved thermal profiles are consistent with a global, monotonically decreasing temperature with altitude, as would be expected from radiative cooling without high altitude absorbers of stellar radiation.  As a test, we compare the retrieved dayside-averaged thermal profile to three scenarios of self-consistent radiative equilibrium models\cite{Fortney2008} and find that it is most congruous with the thermal structure expected at the substellar point (Fig. S\ref{fig:radequilm}).  This result supports our findings of a low day-night heat redistribution.

\begin{figure}
\centering
\if\submitms n
    \includegraphics[width=15cm]{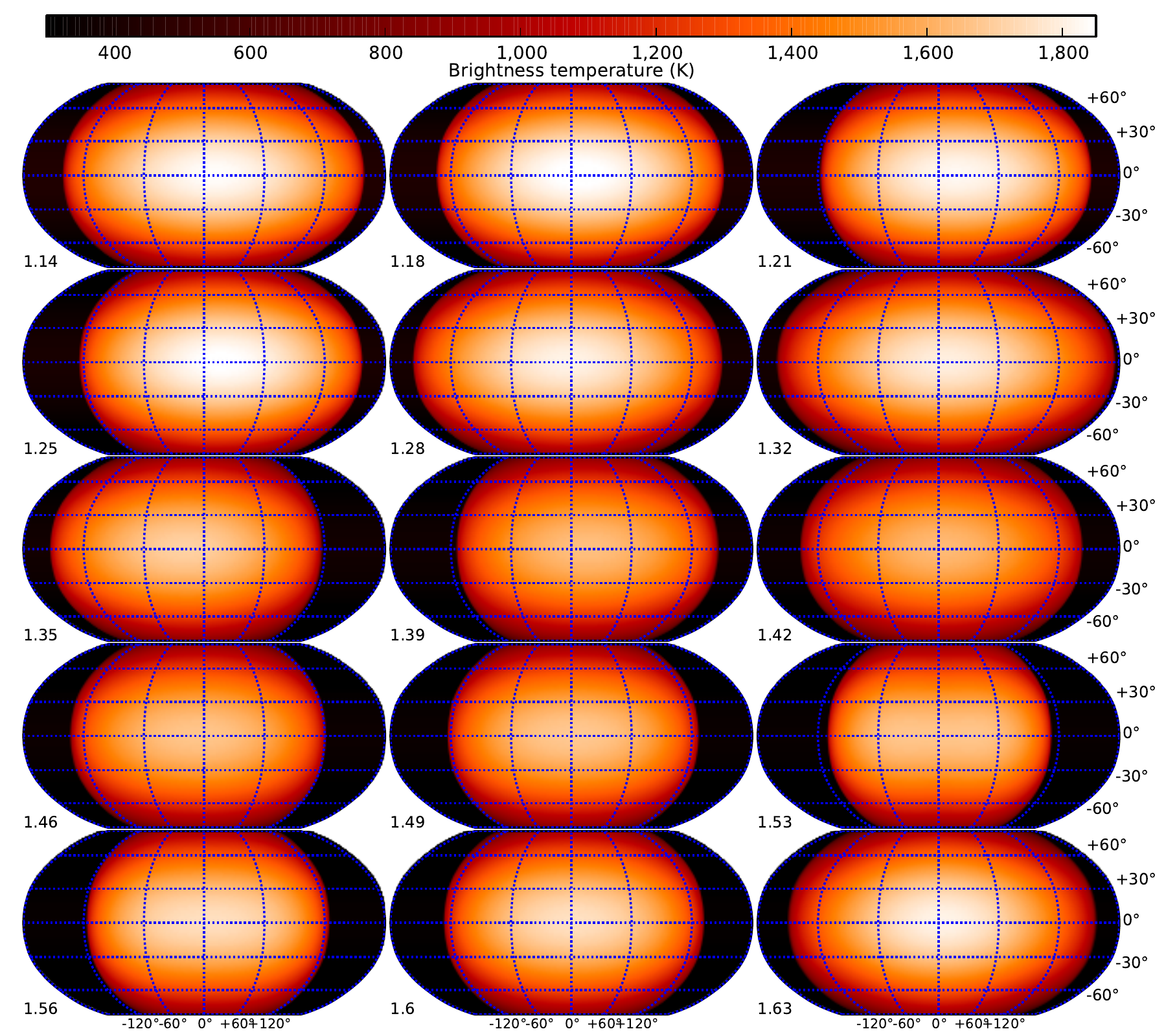}
\fi
\caption{\label{fig:tbmap}
{\bf Longitudinally-resolved brightness temperature maps of WASP-43b in all fifteen spectrophotometric channels.}  Black regions in this Robinson projection indicate no discernible contribution.
Numbers indicate the wavelength in {\microns}.  The observations constrain the brightness temperature at each longitude, but contain no latitudinal information (we assign a $\cos^2$ weighting).  In general, the change in temperature is relatively small over the planet's dayside (-90$^\circ$ to +90$^\circ$) and comparatively extreme near {\pm}120$^\circ$, thus indicating that we detect emission over the planet's entire dayside.  Since WASP-43b does not contain a thermal inversion at these pressures, the hotter regions at a given longitude sample deeper within the atmosphere.  The presence of water vapor in the planet's atmosphere explains the relatively cool brightness temperature from 1.35 -- 1.6 {\microns}.  Outside of the water feature, the brightness temperature peak (indicated in white) is predominantly eastward (towards positive longitudes) of the substellar point.  This correlation is readily seen in Fig. S\ref{fig:pressure+PeakOffset} and matches the predictions of three-dimensional circulation models.  Fig. S\ref{fig:1dltemp} displays one dimensional brightness temperatures with uncertainty regions.
}
\end{figure}

Adopting the same sinusoidal function used to fit the phase variation\cite{SI2014}, we invert the spectroscopic light curves into longitudinally-resolved brightness temperature maps\cite[Fig. \ref{fig:tbmap}]{Cowan2008}.  The brightness temperature, $T_B$, is a function of atmospheric opacity, and water vapor is the main source of opacity in this bandpass.  Because $T_B$ is systematically cooler within the water band, this signifies the global presence of water vapor within the pressure regions probed by these measurements (Fig. S\ref{fig:contributionFunc}). 

The large measured day-night luminosity difference of WASP-43b \cite[$L$\sb{\rm{day}}$/L$\sb{\rm{night}}~$>$~\rm{20 at 1}$\sigma$, \rm{mode} $\sim$~\rm{40}]{SI2014} stands in stark contrast to the modest day-night differences inferred from {\em Spitzer} photometry for giant planets such as HD 189733b, HD 209458b, and HD 149026b that are similarly irradiated \cite{Knutson2007,Knutson2012,Perez-Becker2013}.  Unlike {\em Spitzer} data, our spectrum samples the planet's flux near the peak of its Planck curve, allowing a more robust determination of the total dayside luminosity.  This data set suggests that derived day-night differences may be strongly wavelength dependent and that mid-infrared photometry may not give a complete picture of planetary circulation.

Brightness temperature maps, being functions of both longitude and atmospheric depth, reveal the dynamics of a planet's atmosphere.  Phase-curve peaks prior to the time of secondary eclipse (as seen in Fig. \ref{fig:whitelc}) have previously been reported in hot Jupiters\cite{Knutson2007,Lewis2013} and match predictions from 3D circulation models\cite{Showman2002,CooperShowman2005apjhd209458bweath, Showman2009}.  Such models show that the eastward offset results from a strong jet stream at the equator; our observations thus suggest that WASP-43b exhibits such an eastward-flowing jet.
Our spectrophotometric observations further demonstrate the influence of water vapor on the emergent thermal structure.  Inside the water band (1.35 -- 1.6 {\microns}), observations probe lower atmospheric pressures (higher altitudes) and we measure smaller phase-curve peak offsets relative to the other wavelengths (Figs. S\ref{fig:contributionFunc} and S\ref{fig:pressure+PeakOffset}).  This is qualitatively consistent with variable brown dwarf measurements\cite{Buenzli2012} and circulation-model predictions \cite{CooperShowman2005apjhd209458bweath,Showman2009,Burrows2010,Perez-Becker2013}, which show that smaller displacements are expected at higher altitudes where radiative timescales are much shorter than the relevant dynamical timescales.  However, the observed westward offset of the coldest regions from the antistellar point is puzzling and is not predicted by most models.

The strong day-night temperature variation observed for WASP-43b distinguishes itself from the predominantly uniform temperatures of the Solar System giant planets.  This illustrates the importance of radiative forcing on the atmospheres of close-in exoplanets.  Phase-resolved emission spectroscopy offers a unique way to determine how the extreme stellar radiation incident on these planets is absorbed, circulated, and re-emitted. The door is now open to observations that can constrain theories of planetary atmospheric dynamics in a new regime.

\bibliography{1256758Revisedtext}

\bibliographystyle{Science}

\section*{Acknowledgments}
This work is based on observations made with the NASA/ESA Hubble Space Telescope that were obtained at the Space Telescope Science Institute, which is operated by the Association of Universities for Research in Astronomy, Inc., under NASA contract NAS 5-26555.  Data are available through the Mikulski Archive for Space Telescopes (MAST).
We thank Alison Vick and Merle Reinhart of STScI for scheduling these observations, which are associated with program GO-13467. 
Support for this work was provided by NASA through a grant from the Space Telescope Science Institute, the Sagan Fellowship Program (to K.B.S.) as supported by NASA and administered by the NASA Exoplanet Science Institute (NExScI), the Alfred P. Sloan Foundation through a Sloan Research Fellowship (to J.L.B.), and the National Science Foundation through a Graduate Research Fellowship (to L.K.).
G.W.H. and M.H.W. acknowledge long-term support from NASA, NSF, Tennessee State University, and the State of Tennessee through its Centers of Excellence program.
S.S. acknowledges funding from the Massachusetts Institute of Technology.
D.H. acknowledges support from the European Research Council under the European Community's Seventh Framework Programme -- FP7/2007-2013 Grant Agreement no. 247060.
 

\section*{Supplementary Materials}
Materials and Methods   \\
Supplementary Text      \\
Tables S1 to S2         \\
Figs. S1 to S8          \\
Caption for Movie S1    \\

\clearpage
\renewcommand{\figurename}{{\bf Fig.}}

\renewcommand{\figurename}{{\bf Fig. S}}
\setcounter{figure}{0}
\renewcommand{\tablename}{{\bf Table S}} 
\setcounter{table}{0}

\clearpage
\setcounter{page}{1}
{\centering
\section*{\bf Supplementary Materials for}
{\huge Thermal structure of an exoplanet atmosphere from phase-resolved emission spectroscopy}\\
\vspace*{\baselineskip}

\noindent Kevin B. Stevenson$^{1,13\ast}$, Jean-Michel D\'esert$^2$, Michael R. Line$^{3}$, Jacob L. Bean$^1$,
 Jonathan J. Fortney$^{3}$, Adam P. Showman$^{4}$, Tiffany Kataria$^{4}$, Laura Kreidberg$^1$,
 Peter~R.~McCullough$^{5,6}$, Gregory W. Henry$^{7}$, David Charbonneau$^{8}$, Adam Burrows$^{9}$,
 Sara Seager$^{10}$, Nikku Madhusudhan$^{11}$, Michael H. Williamson$^{7}$ \& Derek Homeier$^{12}$
\\
\vspace*{\baselineskip}

$^\ast$To whom correspondence should be addressed; E-mail:  kbs@uchicago.edu.
\\}
\vspace*{\baselineskip}

{\bf This PDF file includes:}

\indent\indent Materials and Methods\\
\indent\indent Supplementary Text\\
\indent\indent Tables S1 to S2\\
\indent\indent Figures S1 to S8\\
\indent\indent Caption for Movie S1\\

{\bf Other supplementary Materials for this manuscript include the following:}

\indent\indent Movie S1\\
\clearpage

\section*{\bf Materials and Methods}

Each phase curve visit consists of 13 or 14 {\em HST} orbits and each primary transit or secondary eclipse visit consists of four orbits.  To improve observational efficiency, the telescope operated in spatial scan mode, scanning at a rate of 0.08\arcsec s$^{-1}$ and alternating between the forward and reverse directions.  In each scan direction, the instrument made 15 non-destructive reads (SPARS10 sampling) over 103 seconds, the maximum duration possible with these settings.  The observations achieved a duty cycle of $\sim$73\%.  Typically, we acquired 19 exposures per {\em HST} orbit and 1151 exposures total over all visits.  In the extracted 1D spectra, we achieved a signal-to-noise ratio (S/N) of $\sim$1,300 per pixel.  This corresponds to a S/N of $\sim$3,300 per spectrophotometric bin of width 7 pixels.

The WFC3 spatial scan data contain a previously-documented orbit-long systematic that we fit with an exponential ramp model component.  The ramp systematic is steepest during the first {\em HST} orbit and nearly consistent in shape over the remaining orbits.  Some visits have second {\em HST}-orbit ramps that are also noticeably steeper.  Accordingly, we do not include data from the first orbit and, when necessary, fit an additional exponential ramp model component to the second orbit.  Excluding the second orbit from the phase-curve data does not change our conclusions.  

To model each visit-long trend, we use a linear function for the five shorter eclipse/transit observations and a quadratic function for the three longer phase-curve observations.  In each case, we use the Bayesian Information Criterion (BIC) to determine the appropriate order of polynomial.  Using a quadratic trend for the transit/eclipse observations does not change our results.  Using a linear trend for the phase-curve observations results in poor fits in which phase-curve minima fall below the in-eclipse flux (which is physically impossible) for many of the channels.  We also tested multi-visit-long sinusoid models with various periods, but could not achieve better fits than those presented in our final analysis.  We include the curved flux baseline (from the planet's phase variation) in the transit and eclipse models so as to not bias the measured depths.

The sinusoidal function used to represent the band-integrated (white light) phase variation takes the form $c_1\cos[2\pi(t-c_2)/P]+c_3\cos[4\pi(t-c_4)/P]$, where $t$ is time, $P$ is the planet's orbital period, and $c_1$ - $c_4$ are free parameters.  The second sinusoidal term allows us to fit for an asymmetric phase curve, which we detect with $\sim10\sigma$ confidence in the white light curve data.  We do not detect changes in the light-curve due to ellipsoidal variation in the shape of the planet or host star.

In the spectroscopic phase curves, we do not detect statistically significant asymmetry; therefore, we fix $c_3$ and $c_4$ to zero.  Additionally, we fix the ratio between the semi-major axis and the stellar radius ($a/R_{\star}$) and the cosine of the inclination ($\cos i$) in the spectroscopic fits using best-fit values from the white light curve data.  Each spectrophotometric channel shares a common set of eclipse-depth and phase-curve parameters.

We estimate uncertainties using a differential-evolution Markov-chain Monte Carlo (DE-MCMC) algorithm.  Assuming the flux variation is solely from the planet, it is unphysical for the phase-curve to fall below the in-eclipse flux, so we apply an asymmetric prior to $c_1$ (the phase-curve amplitude) wherein credible amplitudes have an uninformative prior and unphysical amplitudes have a Gaussian prior with a standard deviation equal to the eclipse depth uncertainty in each spectrophotometric channel.  

In our analyses of the spectroscopic data, we tested both sinusoid and double-sinusoid models when fitting the phase curves.  We find that the double sinusoid is unjustified according to the BIC.  Nonetheless, we explored the dependence of our free parameters on our choice of model.  Both sets of eclipse depths are consistent to well-within 1$\sigma$.  We also find that the phase-curve amplitudes in 13 of the 15 channels are consistent at the 1$\sigma$ level, and all channels are consistent to within 2$\sigma$.  Although the computed uncertainties in the phase-curve amplitudes and peak offsets for both model combinations are also consistent, we note that four channels exhibit some model dependence ($>2\sigma$) in their best-fit peak offsets.  Relative to the pressure-peak offset trend observed in Fig. S\ref{fig:pressure+PeakOffset}, some outliers with the sinusoid model achieve more consistent peak offsets with the double sinusoid.  However, the latter is also true as some consistent peak offsets with the sinusoid model become outliers with the double sinusoid.  Ultimately, one model combination does not consistently achieve more reliable results than the other.  

\section*{\bf Supplementary Text}

Observations of the thermal emission from the dayside and night side of a planet can inform us on its Bond albedo and heat redistribution efficiency.  Here we derive, using energy balance, the Bond albedo and our metric for estimating the redistribution efficiency.  We need not make use of any reflected-light observations.  First, we must derive the bolometric dayside and night-side fluxes (and their uncertainties) by integrating over wavelength an ensemble of spectra from the MCMC retrieval.  The model spectra are only constrained over the WFC3 bandpass; however, a majority of the flux emanates from near- to mid-infrared wavelengths.  Therefore, we use the MCMC ensemble of fitted atmospheric properties to predict the planetary spectrum out to 20 {\microns}.  This extrapolation contributes to most of the uncertainty in the measured bolometric fluxes.  Upon integrating, we obtain a dayside bolometric flux, $F$\sb{day}, of $(3.9 - 4.1)\times10^{5}$ W m$^{-2}$ and a night-side bolometric flux, $F$\sb{night}, of $< 0.18 \times10^{5}$ W m$^{-2}$ at 1$\sigma$. With these bolometric fluxes, we can compute the desired quantities.   

First, we derive the Bond albedo.  Assuming all of the energy absorbed by the planet is re-radiated and neglecting internal heat from within the planet, we obtain the following relation:
\begin{equation}\label{eq:eb}
S_{\star}(1-A_{\mathrm{B}})\pi R_{\mathrm{p}}^{2}=2\pi R_{\mathrm{p}}^{2}(F_{\mathrm{day}}+F\sb{\mathrm{night}}),
\end{equation} 
where $A$\sb{B} is the Bond albedo and $R$\sb{p} is the planet radius.  The stellar flux at the planet, $S_{\star}$, is given by:
\begin{equation}
S_{\star}=\sigma T_{\star}^{4}\left(\frac{R_{\star}}{a}\right)^{2},
\end{equation} 
where $\sigma$ is the Steffan-Boltzman constant, $T_{\star}$ is the stellar effective temperature, $R_{\star}$ is the stellar radius, and $a$ is the planet's semi major axis.  The left-hand-side (LHS) of Equation \ref{eq:eb} is the stellar flux incident upon the planet and the right-hand-side (RHS) is the flux re-radiated from the planet.   Using our computed  $F$\sb{day} and $F$\sb{night} values, the measured stellar effective temperature (4,520~$\pm$~120 K), and the measured $a/R_{\star}$ (4.855~$\pm$~0.002), we determine the Bond albedo to be 0.078 -- 0.262.

Second, we rewrite the heat redistribution efficiency in terms of our observed quantities.  If both planet sides have the same temperature ($F$\sb{day}$=F$\sb{night}, full redistribution) then Equation \ref{eq:eb} becomes:
\begin{equation}\label{eq:eb2}
S_{\star}(1-A_{\mathrm{B}})\pi R_{\mathrm{p}}^{2}=4\pi R_{\mathrm{p}}^{2}F_{\mathrm{day}}\mathcal{F},
\end{equation} 
where $\mathcal{F}$ is the redistribution factor, which is unity in the case of full redistribution.  We equate the RHS of Equation \ref{eq:eb} to the RHS of Equation \ref{eq:eb2} and then solve for the redistribution factor:
\begin{equation}\label{eq:redist2}
\mathcal{F}=\frac{1}{2}(1+\frac{F_{night}}{F_{day}}).
\end{equation}
In the case of full redistribution ($F$\sb{day}$=F$\sb{night}), we recover $\mathcal{F}=1$.  If there is no redistribution, meaning all of the flux emanates from only the dayside ($F$\sb{night}$=0$), then $\mathcal{F}=0.5$.  Inputing the measured $F$\sb{day} and $F$\sb{night} values, we find that $\mathcal{F}=$ 0.500 -- 0.524.

\begin{table}
\centering
\captionsetup{font=footnotesize}
\caption{\label{tab:WhiteData}
{\bf Best-Fit White Light Parameters with 1$\sigma$ Uncertainties}}
\footnotesize   
\begin{tabular}{lllll}
    \hline
    Parameter                   & Value   \\
    \hline
    Transit Times (BJD\sb{TDB}) & 2456601.02729(2)  \\
                                & 2456602.65444(2)  \\
                                & 2456603.46792(2)  \\
                                & 2456605.90822(2)  \\
                                & 2456612.41604(3)  \\
                                & 2456615.66978(1)  \\
    $R$\sb{p}$/R$\sb{$\star$}   & 0.15948(4)        \\
    $a/R$\sb{$\star$}           & 4.855(2)          \\
    $\cos i$                    & 0.13727(19)       \\
    Eclipse Times (BJD\sb{TDB}) & 2456601.43503(16) \\
                                & 2456602.25412(14) \\
                                & 2456603.87485(13) \\
                                & 2456608.75729(23) \\
                                & 2456632.34584(12) \\
    Eclipse Depth (ppm)         & 461(5)            \\
    $c$\sb{1} (ppm)             & 234(2)            \\
    $c$\sb{2} (BJD\sb{TDB})     & 2456601.4290(12)  \\
    $c$\sb{3} (ppm)             & 29(1)             \\
    $c$\sb{4} (BJD\sb{TDB})     & 2456601.3486(15)  \\
    \hline
\end{tabular}
\\
\begin{minipage}{0.41\textwidth}\footnotesize
BJD\sb{TDB}, Barycentric Julian Date, Barycentric Dynamical Time; ppm, parts per million.  Parentheses indicated 1$\sigma$ uncertainties in the least significant digit(s).
\end{minipage}
\end{table}

\begin{table}
\centering
\captionsetup{font=footnotesize}
\caption{\label{tab:SpecData}
{\bf Best-Fit Spectroscopic Parameters with 1$\sigma$ Uncertainties}}
\footnotesize
\begin{tabular}{lllll}
    \hline
    Wavelength      & PC amplitude  & PC peak offset    & Eclipse depth & Dayside $T_{B}$   \\
    ({\microns})    & (ppm)         & (minutes)         & (ppm)         & (K)               \\
    \hline
             1.1425 &    177{\pm}16 &        -28{\pm}32 &    367{\pm}45 &   1,809{\pm}33 \\
             1.1775 &    213{\pm}13 &        -28{\pm}24 &    431{\pm}39 &   1,826{\pm}25 \\
             1.2125 &    215{\pm}13 &        -57{\pm}24 &    414{\pm}38 &   1,791{\pm}25 \\
             1.2475 &    242{\pm}12 &        -51{\pm}20 &    482{\pm}36 &   1,814{\pm}21 \\
             1.2825 &    216{\pm}15 &         12{\pm}18 &    460{\pm}37 &   1,778{\pm}23 \\
             1.3175 &    212{\pm}17 &        -26{\pm}21 &    473{\pm}33 &   1,765{\pm}20 \\
             1.3525 &    186{\pm}10 &         63{\pm}26 &    353{\pm}34 &   1,669{\pm}26 \\
             1.3875 &    167{\pm}10 &        -51{\pm}26 &    313{\pm}30 &   1,620{\pm}25 \\
             1.4225 &    162{\pm}11 &        -11{\pm}21 &    320{\pm}36 &   1,607{\pm}30 \\
             1.4575 &    206{\pm} 7 &         23{\pm}13 &    394{\pm}36 &   1,646{\pm}26 \\
             1.4925 &    228{\pm} 9 &         -6{\pm}17 &    439{\pm}33 &   1,657{\pm}22 \\
             1.5275 &    244{\pm} 5 &         -3{\pm}17 &    458{\pm}35 &   1,664{\pm}23 \\
             1.5625 &    306{\pm} 8 &         -8{\pm}11 &    595{\pm}36 &   1,728{\pm}20 \\
             1.5975 &    309{\pm}12 &         -9{\pm}12 &    614{\pm}37 &   1,723{\pm}20 \\
             1.6325 &    344{\pm}17 &        -12{\pm}12 &    732{\pm}42 &   1,772{\pm}20 \\
    \hline
\end{tabular}
\begin{minipage}{0.7\textwidth}\footnotesize
PC, phase curve;  ppm, parts per million.  The peak offset is with respect to the fixed time of mid-eclipse, as determined from a white-light-curve fit.  We use a 4,520 K stellar Kurucz model when estimating the dayside brightness temperatures ($T$\sb{B}).
\end{minipage}
\end{table}
           
\clearpage

\begin{figure}
\centering
\if\submitms n
    \includegraphics[width=15cm]{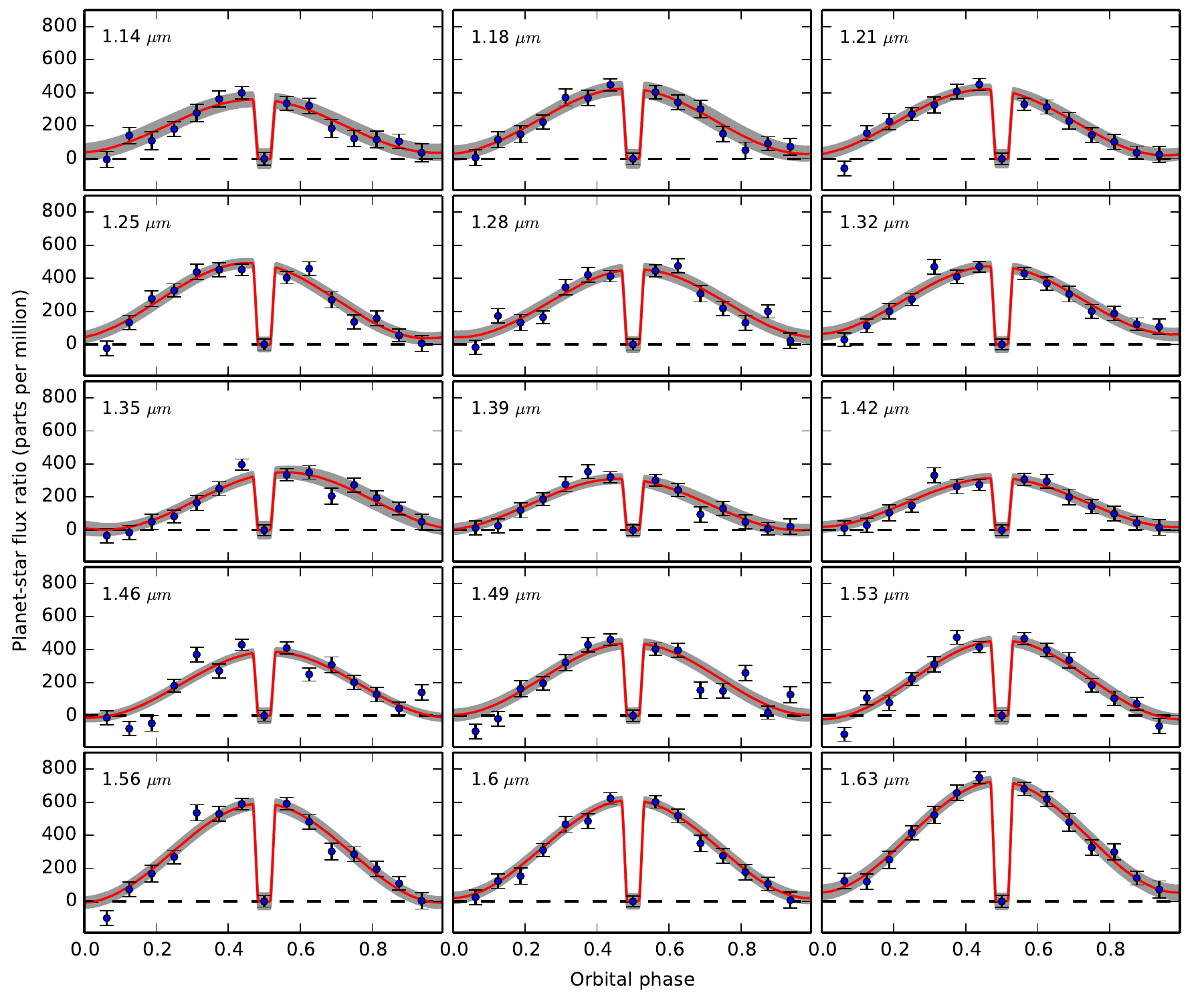}
\fi
\caption{\label{fig:PCall}
{\bf Spectroscopic phase curves of WASP-43b.}
Red curves indicate median models to the blue data points with 1$\sigma$ uncertainties.  The gray regions indicate model 1$\sigma$ uncertainty regions.
}
\end{figure}

\begin{figure}
\centering
\if\submitms n
    \includegraphics[width=15cm]{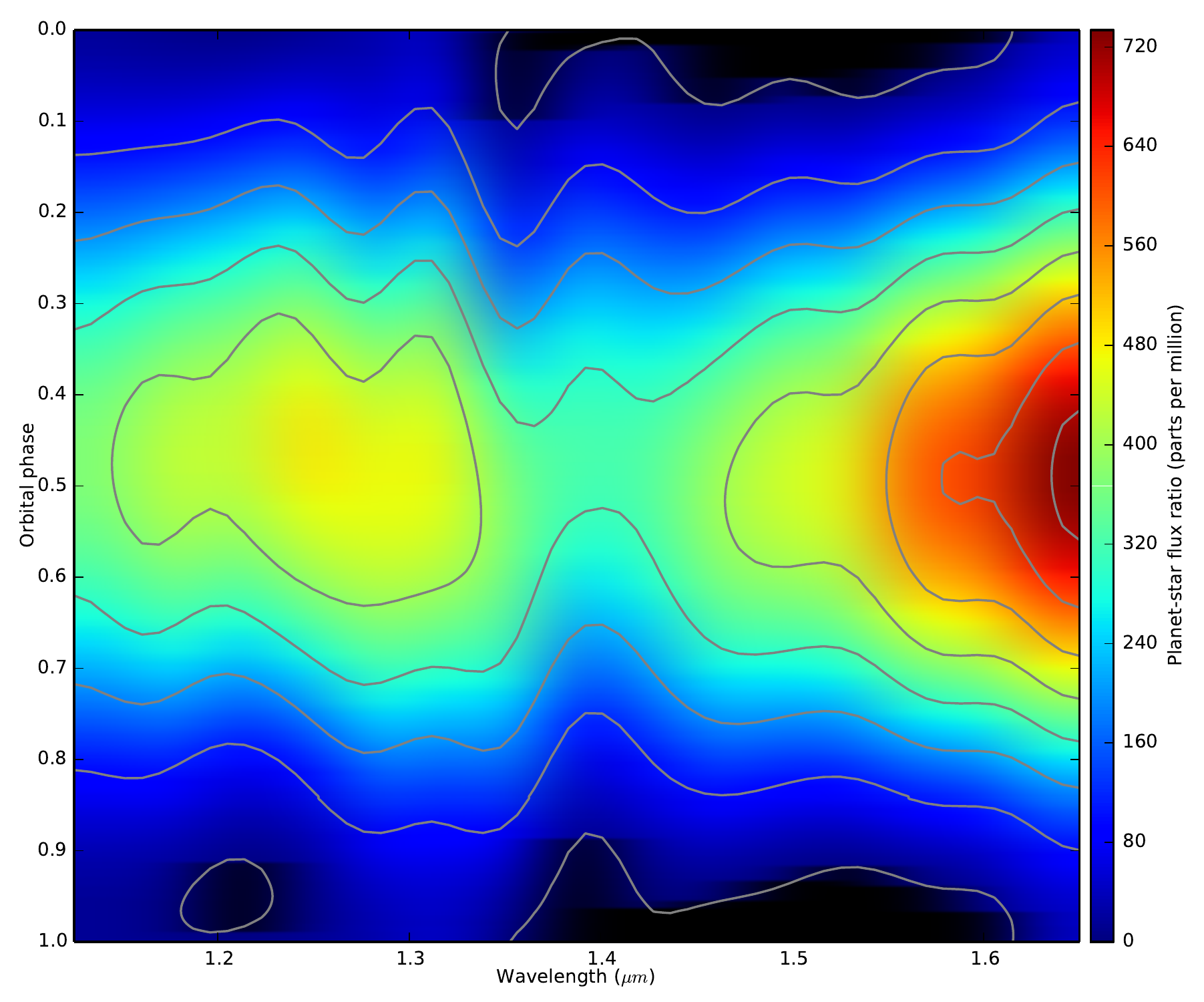}
\fi
\caption{\label{fig:2DEmSpec}
{\bf Phase-resolved emission spectrum of WASP-43b relative to the stellar flux.}  
To generate this map, we apply bi-cubic interpolation between our 15 best-fit spectroscopic phase curve models.  The eight contour lines are evenly spaced from minimum to maximum planet emission.
}
\end{figure}

\begin{figure}
\centering
\if\submitms n
    \includegraphics[width=15cm]{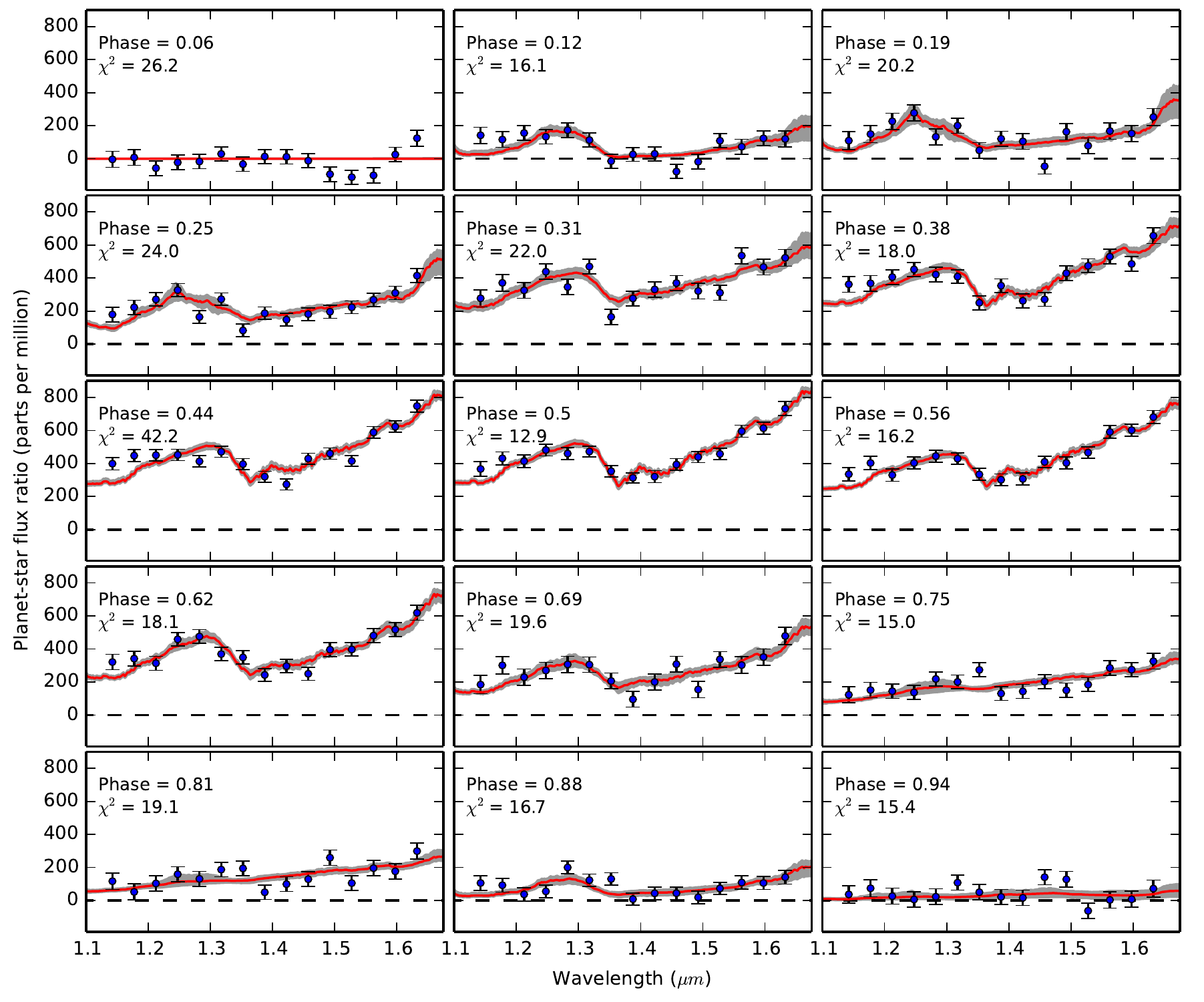}
\fi
\caption{\label{fig:EmSpecall}
{\bf Emission spectra of WASP-43b at fifteen binned orbital phases.}
Although the data are subdivided into sixteen bins, the last bin occurs during transit (phase = 0.0), when we have no information about the planet's thermal emission.  Red lines indicate median models to the blue data points with 1$\sigma$ uncertainties.  The gray regions indicate model 1$\sigma$ uncertainty regions. 
}
\end{figure}

\begin{figure}
\centering
\if\submitms n
    \includegraphics[width=15cm]{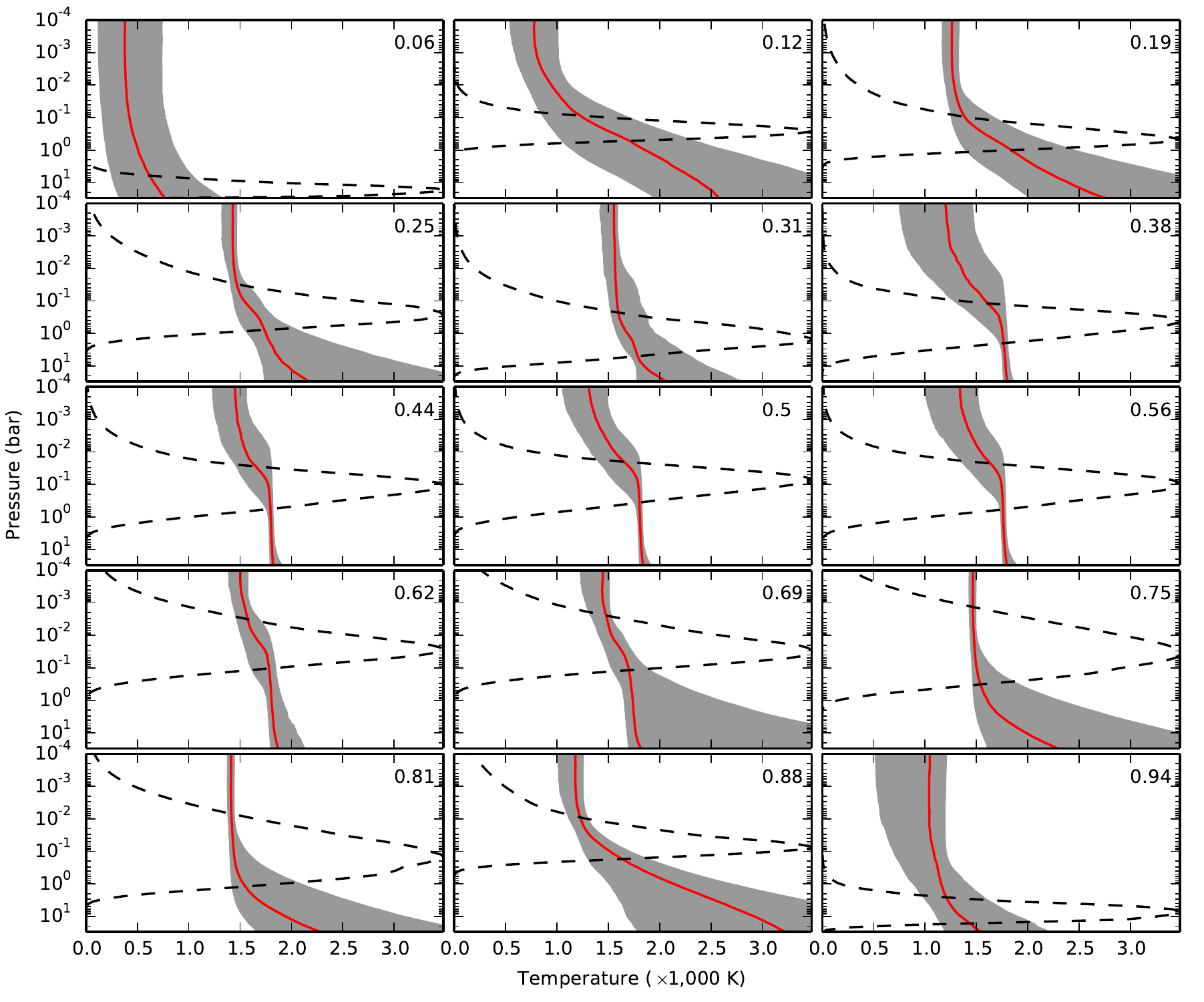}
\fi
\caption{\label{fig:tpall}
{\bf Thermal profiles of WASP-43b at fifteen binned orbital phases.}
Red curves depict median thermal profiles with gray 1$\sigma$ uncertainty regions for the assumed parameterization of the retrieval.  The dashed black curves are the WFC3 bandpass-averaged thermal emission contribution functions at each orbital phase.  These contribution functions illustrate the atmospheric depths at which the observations probe.  Therefore, the temperature retrieval results are most reliable within the pressure levels encompassed by the contribution functions.  We infer temperatures outside of these regions based on the thermal profile parameterization and not by any explicit use of priors.
}
\end{figure}

\begin{figure}
\centering
\if\submitms n
    \includegraphics[width=15cm]{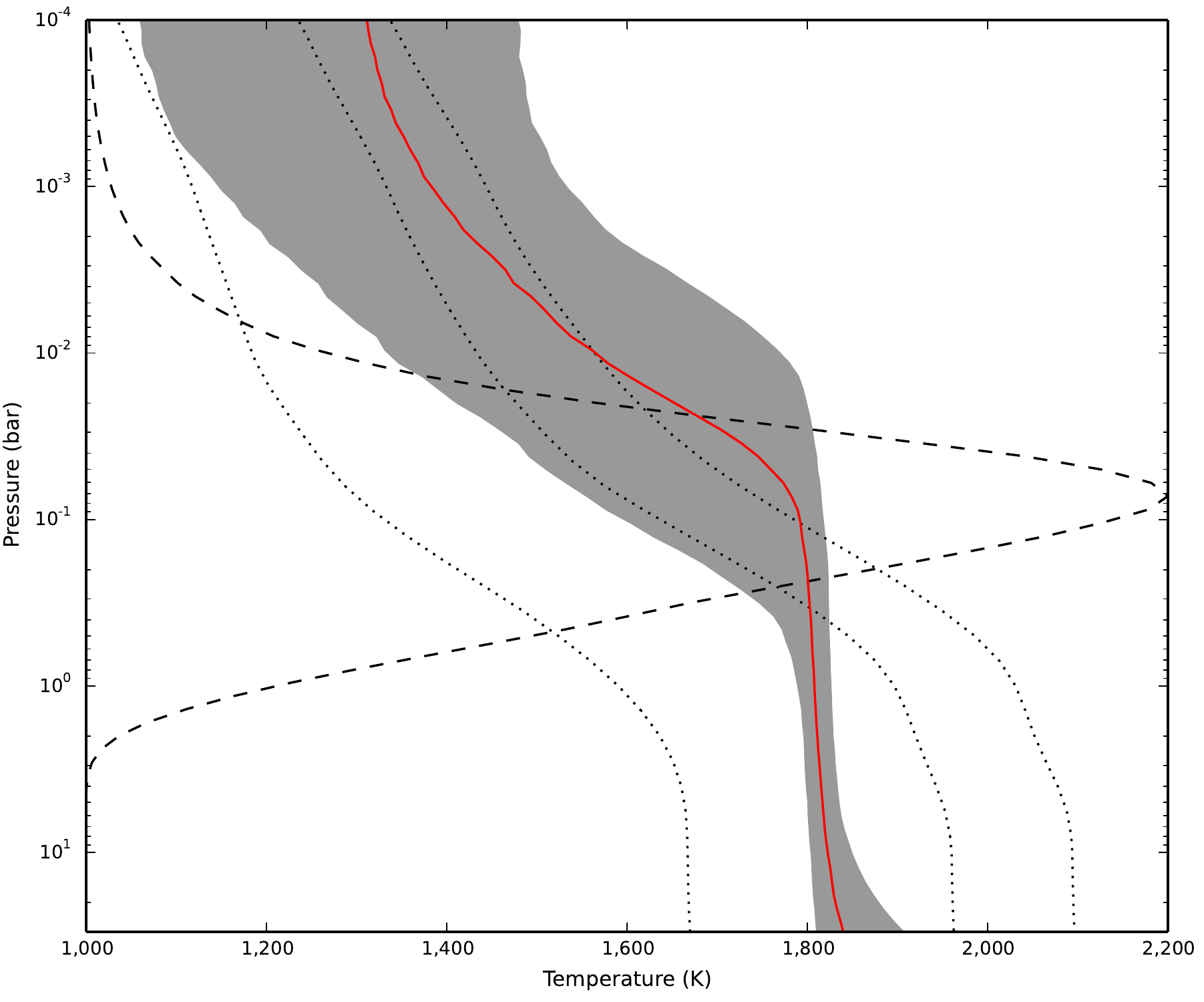}
\fi
\caption{\label{fig:radequilm}
{\bf A comparison between the retrieved dayside and self-consistent temperature profiles.} 
The solid red curve and gray region represent the median and 1$\sigma$ uncertainty limits of the retrieved temperature profile from the WASP-43b secondary-eclipse data.  The dashed black curve is the averaged thermal emission contribution function over the WFC3 bandpass.  The dotted black curves are the temperature profiles computed from a self-consistent radiative equilibrium model\cite{Fortney2008}.  They represent, from cool to hot respectively, 4$\pi$ (full planet) heat redistribution, 2$\pi$ (dayside only) heat redistribution, and the substellar point.  The retrieved thermal profile is consistent with the latter two radiative equilibrium models over the regions probed by these observations and best fits the self-consistent temperature profile at the substellar point.  This suggests that the retrieval is heavily weighted towards fluxes from the substellar point and that the planet's day-night heat redistribution is inefficient, in accordance with the phase curve.
}
\end{figure}

\begin{figure}
\centering
\if\submitms n
    \includegraphics[width=15cm]{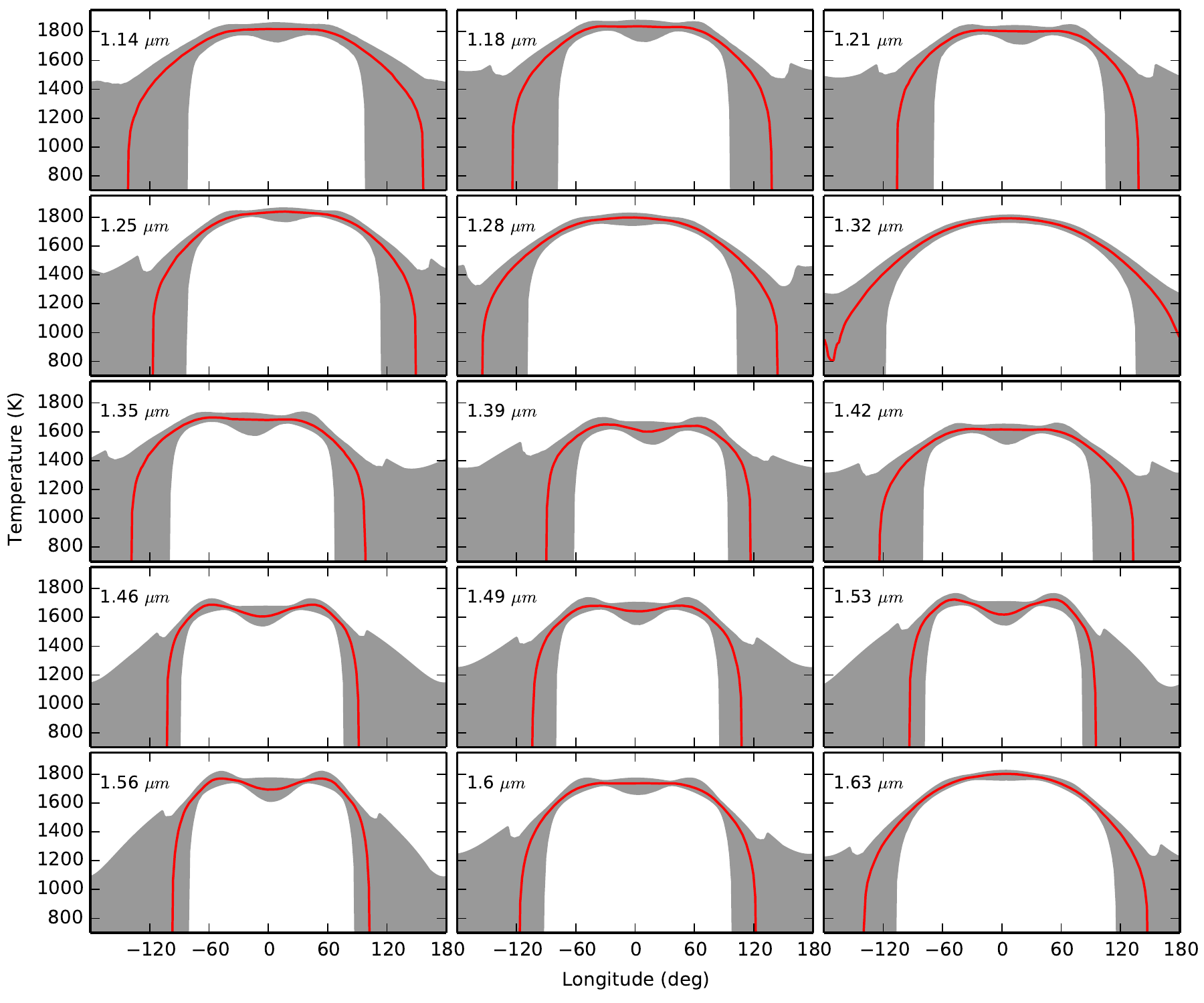}
\fi
\caption{\label{fig:1dltemp}
{\bf Longitudinally-resolved brightness temperatures at fifteen spectrophotometric channels.} 
Red lines indicate median models and gray regions depict 1$\sigma$ uncertainty regions.  We generate these models by inverting 1,000 phase curves per spectroscopic channel from our DEMCMC analysis into longitudinally-dependent light curves by way of a least-squares minimizer, computing the brightness temperatures for each model at every longitude, and then estimating the asymmetric uncertainty regions about the median.  Near the night side, all of the uncertainties extend down to 0 K, thus indicating no lower constraints on the night-side temperatures.  The apparent dips in the median fits near the expected peak hotspots ($\sim$0$^{\circ}$) in some channels is a byproduct of our five-parameter sinusoidal model parameterization and is not physical.  Within the water band, the strong dayside emission in combination with the absence of measured flux on the planet's night-side create a steep temperature gradient near the terminator.  Our primary-mode (lower-frequency) sinusoid achieves a good fit to the steep gradient but over-predicts the dayside temperature plateau.  The secondary-mode (higher-frequency) sinusoid negates the primary at its peak to create a flattened model.  Over-dampening at the peaks is what causes these apparent dips.  With additional terms we can achieve more realisitic fits; however, according to the BIC, the quality of the data does not warrant more complex sinusoidal models.
}
\end{figure}

\begin{figure}
\centering
\if\submitms n
    \includegraphics[width=15cm]{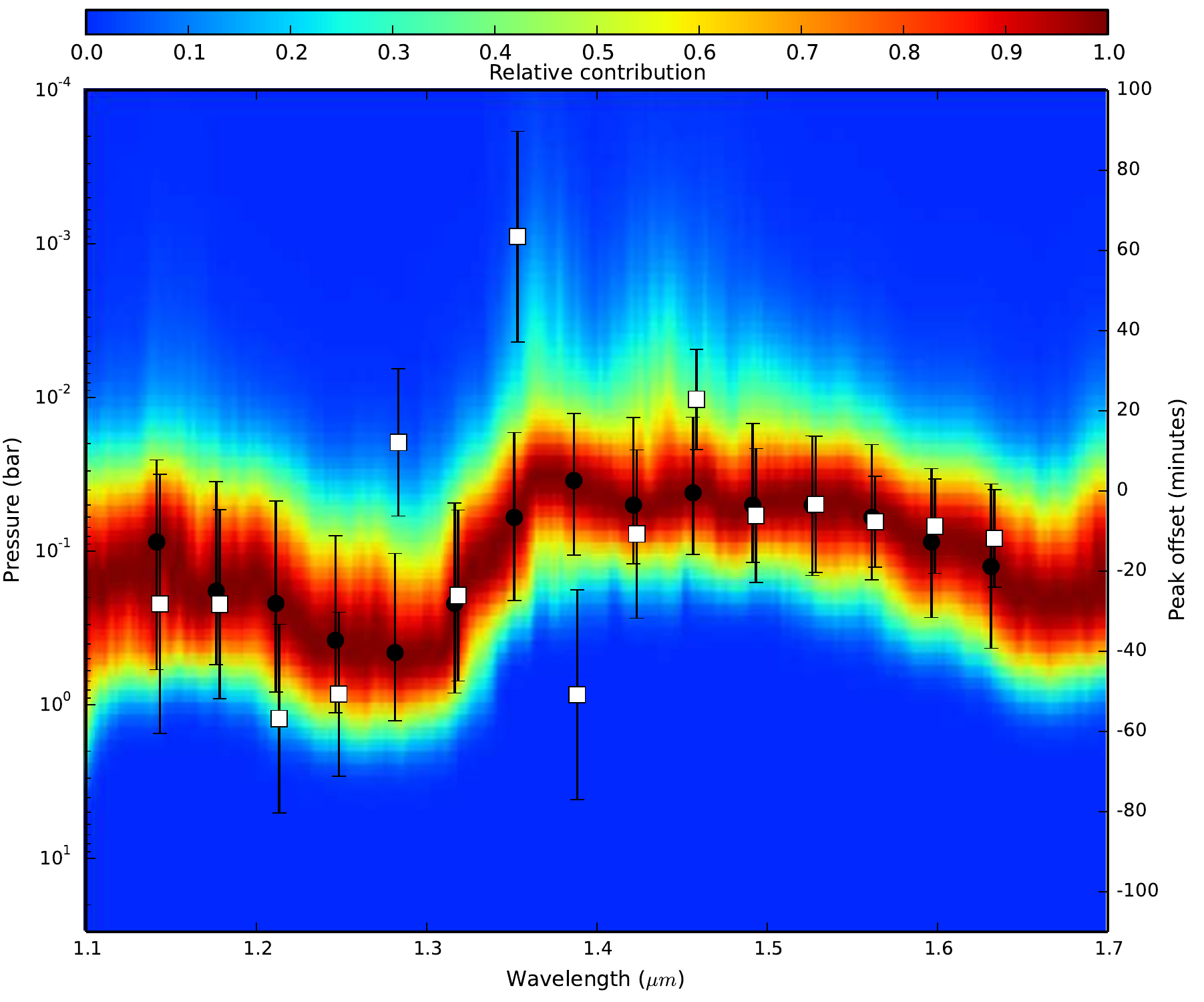}
\fi
\caption{\label{fig:contributionFunc}
{\bf Dayside thermal emission contribution function of WASP-43b.}
The function is computed using the median values from the secondary-eclipse retrieval.  Red indicates the pressure levels at which the optical depth is unity.  These regions have the most significant contribution to the wavelength-dependent emission.  Blue indicates regions with negligible contribution to the total emission.  At high pressures, the column of gas is too dense for the slant rays to penetrate and, at low pressures, the column density is too low for the gas to significantly impact the spectrum.  Black circles signify the pressure level at peak contribution in each spectrophotometric channel and vertical lines represent the full-width at half maximum.  The dayside emission emanates primarily between 0.01 and 1 bar.  White squares with 1$\sigma$ uncertainties represent the phase-curve peak offsets from Table S\ref{tab:SpecData} (scaling on right axis).  Despite the outliers, there is a visible correlation between the dayside thermal emission contribution levels and phase-curve peak offsets as a function of wavelength.
}
\end{figure}

\begin{figure}
\centering
\if\submitms n
    \includegraphics[width=15cm]{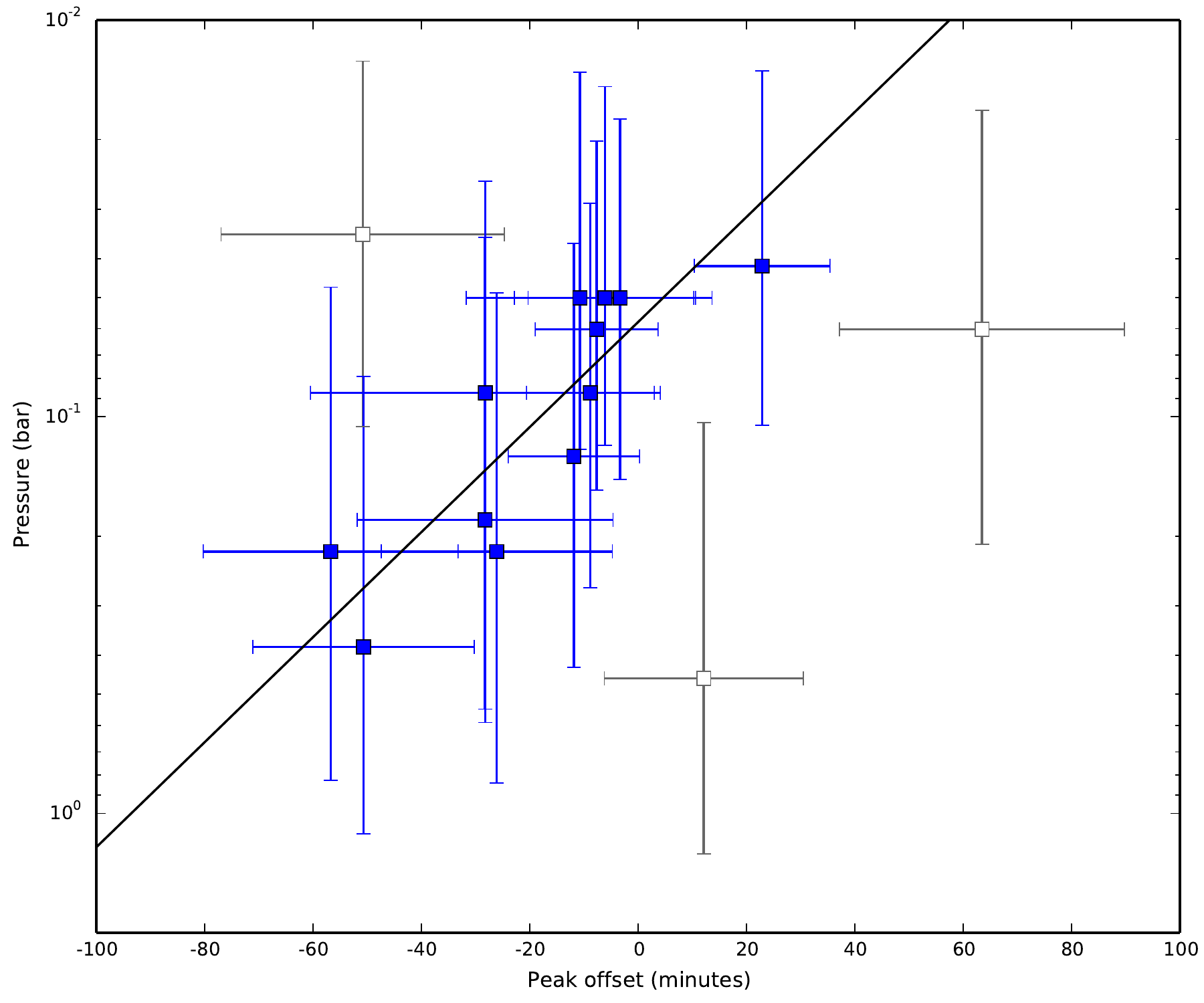}
\fi
\caption{\label{fig:pressure+PeakOffset}
{\bf  Correlation between the dayside thermal emission contribution level and phase-curve peak offset.}
The vertical error bars represent the full-width at half maxima from Fig. S\ref{fig:contributionFunc} and the horizontal error bars are 1$\sigma$ peak offset uncertainties from Table S\ref{tab:SpecData}.  We use orthogonal distance regression to fit a linear model (black line) to the 12 good channels (blue squares) and measure a slope with a significance of 5.6$\sigma$ with respect to the null hypothesis (no correlation).  These data have a Pearson product-moment correlation coefficient of -0.85, where 0 is no correlation and -1 is total inverse correlation.  We use Chauvenet's criterion on the measured peak offsets to identify the three white squares as outliers.  This can be seen in Fig. S\ref{fig:contributionFunc} where the 1.2825, 1.3525, and 1.3875 {\micron} channels do not have peak offsets that vary smoothly with wavelength as influenced by the broad water feature.  At higher atmospheric pressures, we detect a stronger deviation in the phase-curve peak offset relative to the time of secondary eclipse.  This trend qualitatively matches the predictions of three-dimensional circulation models.
}
\end{figure}

\setcounter{figure}{0}
\renewcommand{\figurename}{{\bf Movie S}} 


\centering
\if\submitms n
    \includemovie[text={\includegraphics[width=15cm]{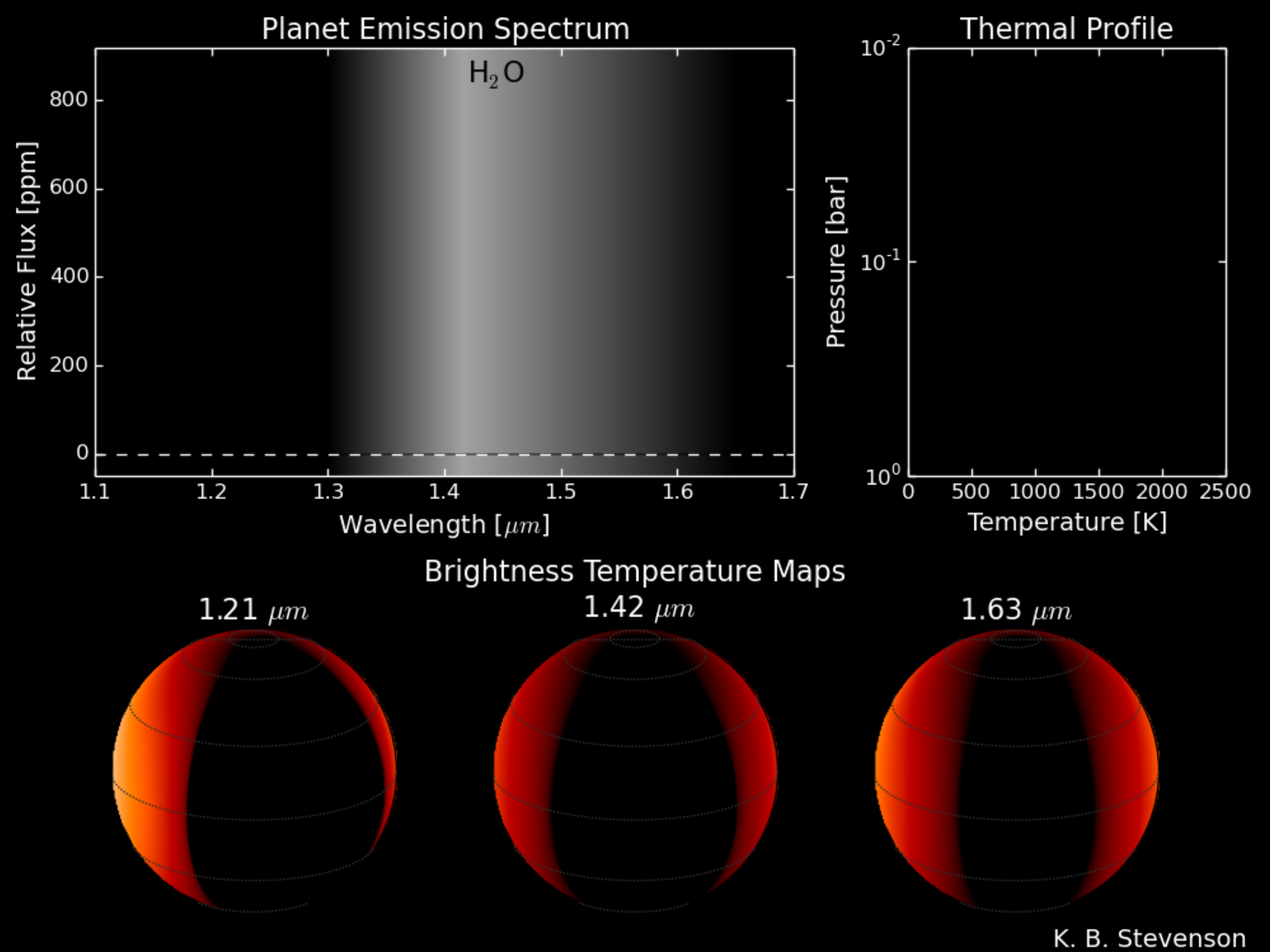}}]{}{}{Stevenson+Etal-2014-WASP43b-Python-m1.mp4}
\fi
\begin{figure}[h!]
\caption{{\bf  Time-lapse video of WASP-43b over one planet rotation.}
The left and right panels display the phase-resolved emission spectrum and thermal profile, respectively, with 1$\sigma$ uncertainty regions.  There is a broad water absorption feature from 1.35 to 1.6 {\microns}.  The rotating spheres depict longitudinally-resolved brightness temperature maps in three spectrophotometric channels.  The video is also available at \href{http://astro.uchicago.edu/~kbs/wasp43b.html}{\color{blue}http://astro.uchicago.edu/$\sim$kbs/wasp43b.html}.
}
\end{figure}

\end{document}